\documentclass[superscriptaddress,longbibliography]{revtex4}

\pdfoutput=1

\usepackage[OT1]{fontenc}
\usepackage[english]{babel}
\usepackage{graphicx}
\usepackage{color}
\usepackage{amssymb,amsmath}
\usepackage{amssymb}
\usepackage[Gray,squaren]{SIunits}
\usepackage{xspace}
\usepackage{upgreek}
\usepackage{ulem}
\usepackage{hyperref}
\usepackage{bm}
\usepackage{notes2bib}
\newcommand{\ud}{\mathrm{d}}

\newcommand{\ie}{i.e.\@\xspace}
\newcommand{\eg}{e.g.\@\xspace}

\newcommand{\eq}[1]{Eq.~\eqref{eq:#1}}
\newcommand{\eqs}[2]{Eqs.~\eqref{eq:#1} and~\eqref{eq:#2}}

\newcommand{\fig}[1]{Fig.~\ref{fig:#1}}

\newcommand{\erfc}{\operatorname{erfc}}
\newcommand{\im}{\operatorname{Im}}
\newcommand{\re}{\operatorname{Re}}

\newcommand{\vbar}{\bar{v}}

\addto\captionsenglish{}

\begin{document}

\title{Large optical depth frequency modulation spectroscopy}

\author{C. C. Kwong}
\email{kwon0009@e.ntu.edu.sg}
\affiliation{School of Physical and Mathematical Sciences, Nanyang Technological University, 637371 Singapore, Singapore}
\affiliation{MajuLab, International Joint Research Unit UMI 3654, CNRS, Universit\'{e} C\^{o}te d'Azur, Sorbonne Universit\'{e}, National University of Singapore, Nanyang Technological University, Singapore}

\author{E. A. Chan}
\affiliation{School of Physical and Mathematical Sciences, Nanyang Technological University, 637371 Singapore, Singapore}
\affiliation{Centre for Disruptive Photonic Technologies, Nanyang Technological University, 637371 Singapore}

\author{S. A. Aljunid}
\affiliation{Centre for Disruptive Photonic Technologies, Nanyang Technological University, 637371 Singapore}

\author{R. Shakhmuratov}
\affiliation{Kazan Physical-Technical Institute, FRC Kazan Scientific Center of RAS, Kazan 420029 Russia}
\affiliation{Kazan Federal University, Kazan 420018 Russia}

\author{D. Wilkowski}
\affiliation{School of Physical and Mathematical Sciences, Nanyang Technological University, 637371 Singapore, Singapore}
\affiliation{MajuLab, International Joint Research Unit UMI 3654, CNRS, Universit\'{e} C\^{o}te d'Azur, Sorbonne Universit\'{e}, National University of Singapore, Nanyang Technological University, Singapore}
\affiliation{Centre for Disruptive Photonic Technologies, Nanyang Technological University, 637371 Singapore}
\affiliation{Centre for Quantum Technologies, National University of Singapore, 117543 Singapore, Singapore}

%
%




\begin{abstract}
Band-resolved frequency modulation spectroscopy is a common method to measure weak signals of radiative ensembles. When the optical depth of the medium is large, the signal drops exponentially and the technique becomes ineffective. In this situation, we show that a signal can be recovered when a larger modulation index is applied. Noticeably, this signal can be dominated by the natural linewidth of the resonance, regardless of the presence of inhomogeneous line broadening. We implement this technique on a cesium vapor, and then explore its main spectroscopic features. This work opens the road towards measurement of cooperative emission effects in bulk atomic ensemble. 
\end{abstract}

\maketitle

\section{Introduction}
Band-resolved frequency modulation (FM) spectroscopy was proposed in 1980 by G. Bjorklund, as a sensitive method to measure absorption and dispersion of weak transmission signals \cite{Bjorklund:80}. Here, the carrier frequency of a laser is scanned across the resonance of the transition under investigation. In the weak modulation index limit, the  amplitude and phase modification of the carrier component are encoded in the beat note with the first sidebands. Band-resolved FM spectroscopy and its variants like the Pound-Drever-Hall technique \cite{drever1983laser, black2001introduction}, or the modulation transfer spectroscopy \cite{PhysRevLett.44.1251,doi:10.1063/1.92867,shirley1982modulation} are key laser spectroscopic techniques for numerous applications such as laser frequency stabilization \cite{mccarron2008modulation,zi2017laser}, Doppler-free spectroscopy \cite{bjorklund1981sub,sansonetti1995measurements,ye1996sub},  detection of gases \cite{silver1992frequency,ma1999ultrasensitive,cygan2011active,goldenstein2014wavelength,sur2015scanned}, magnetometers \cite{budker2002nonlinear} and strain sensors~\cite{liu2011ultra,gatti2008fiber}.

At large optical depth (OD), the carrier is strongly absorbed and the usual transmission FM spectroscopy method is ineffective. Thus, FM spectroscopy measurements on strongly absorbing media are usually performed using thin penetration layers, such as in selective reflection spectroscopy \cite{woerdman1975spectral}, where measurements of collisional broadening  \cite{akulshin1982,maki1991} and atom-surface interaction have been reported \cite{oria1991spectral,chan2018tailoring}. In addition, cooperative atomic emissions have been investigated in dense atomic media, using both cold atomic gases \cite{kwong2014cooperative,pellegrino2014,PhysRevLett.117.073003,guerin2016subradiance,jennewein2016,PhysRevA.96.053629,saint2018resonant} and hot atomic vapors \cite{keaveney2012,silans2018,PhysRevLett.120.243401}. In the latter, large absorption of the transmitted signal is avoided using nano-cells~\cite{sarkisyan2001sub}. However, it is challenging to discern between the bulk cooperative properties and finite-size effects coming from atom-surface interactions~\cite{Fichet_2007}, non-Maxwellian velocity distributions \cite{todorov2017testing} or Dicke-like narrowing \cite{dutier2003collapse}.

In this article, we explore a new FM spectroscopic method that has a good sensitivity when applied on a medium with large OD. We perform FM at large modulation index to suppress the strongly absorbed on-resonance carrier component. As a result, the on-resonance signal is dominated by the weakly absorbed sidebands, which probe the tails of the resonance dominated by the slow algebraic decay of the homogeneous linewidth, rather than the faster exponential decay of some frequency broadening mechanisms (\eg  Doppler effect). For a large OD medium, we show that the frequency sensitivity of this technique is comparable to the standard FM spectroscopy at low temperature. Importantly at high temperature, the sensitivity of the new method remains unchanged because it is not affected by Doppler broadening.

\section{Experimental study}
\subsection{Experimental setup and parameters}
\begin{figure}[htbp]
   \begin{center}
      \includegraphics[width =\linewidth]{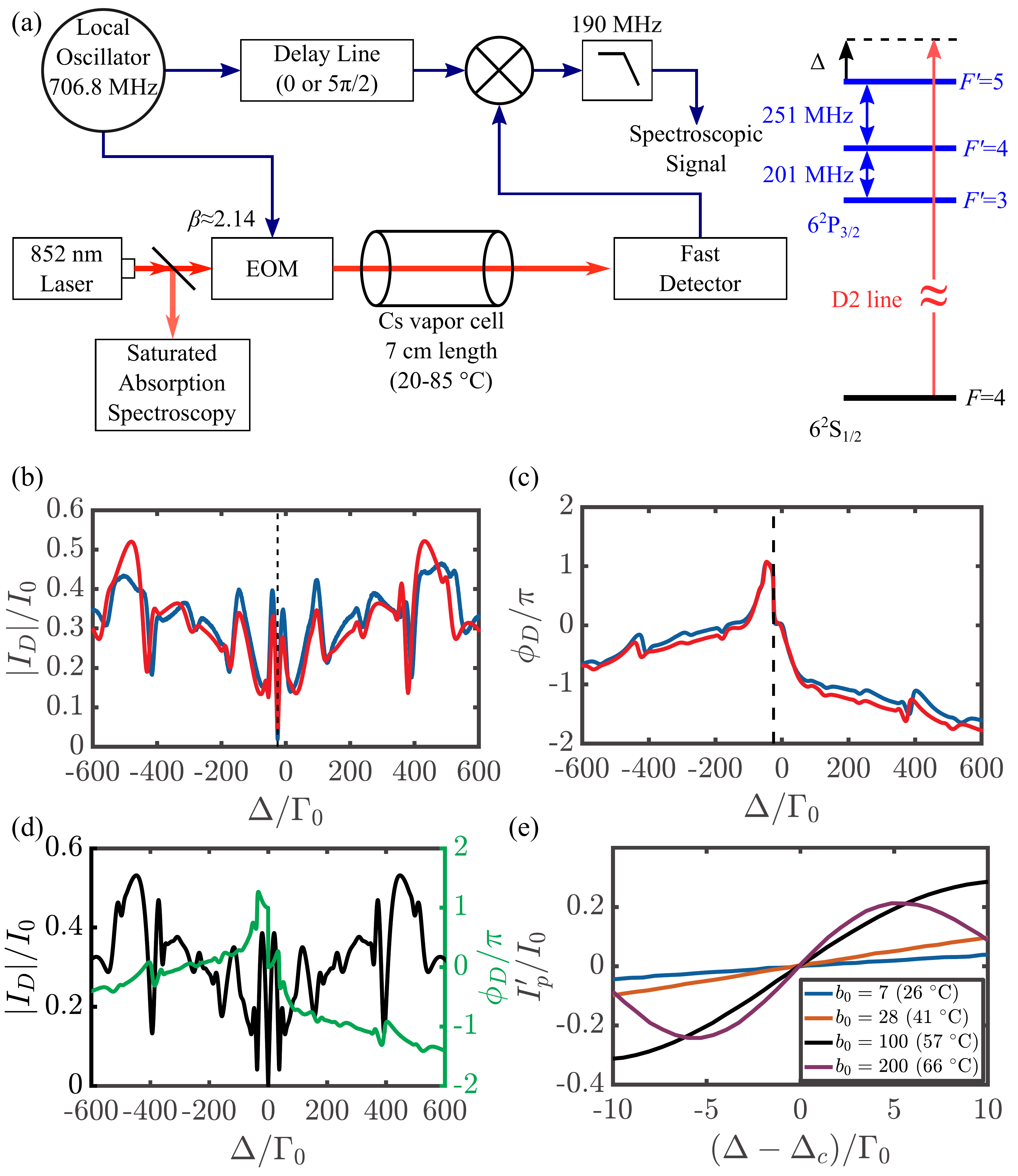}
      \caption{(a) The experimental setup and the relevant cesium energy level. (b)-(c) A comparison between experimental results (blue curve) and theoretical predictions (red curve) for the  magnitude and the phase of the demodulated signals at $T=53$~$^\circ$C. The detuning $\Delta$ is measured from the $F=4\rightarrow F'=5$ transition. In (b), the minimum of the demodulated signal, indicated by the black dashed line, is identified as the spectrum center $\Delta_c$. In (c), the center occurs at the black dashed line, when $\phi_D$ rapidly changes by $\pi$.  (d) The  theoretically calculated  magnitude (black curve) and the phase (green curve) of the demodulated signal for a two-level medium at the same density and temperature with (b) and (c). (e) A zoom around $\Delta_c$ showing the $I'_p$ component of the demodulated signal, which has the full slope of the demodulated signal at $\Delta_c$ (see text for more details). $I_0$ is the incident laser intensity.}
      \label{fig:1}
   \end{center}
\end{figure}

The experiment is performed as follow: A 852 nm laser is scanned across the $F=4\rightarrow F'=3,4,5$ hyperfine transitions of the cesium D2 line (natural linewidth: $\Gamma/2\pi=5.2\,$MHz). The optical frequency is calibrated on a standard saturated absorption spectroscopy setup. The laser beam is sent on a single passage to another $L=7$~cm long cesium vapor cell, heated to a temperature in the range of 20--85~$^\circ$C, resulting in an OD in the range of $b_0=\,$3--700 [see \fig{1}(a)].  A local oscillator of frequency $\Omega=2\pi\times 706.8\,\textrm{MHz}=135.9\Gamma$, generated by a voltage controlled oscillator of maximum frequency 750~MHz, drives an electro-optic modulator (EOM) to generate the phase modulation with a large modulation index of $\beta=2.14(10)$. Using a fast detector, a mixer, and a low-pass filter, the transmitted signal is demodulated at the reference  frequency $\Omega$. With a fixed delay line of 5$\pi$/2, we extract the full demodulated signal $I_D=I_P+iI_Q$, where $I_P$ and $I_Q$ are in-phase and in-quadrature components, respectively (see \ref{sec:generaltheory} for a theoretical description of these components). We used an amplitude modulated signal to calibrate the overall transfer gain of our detection scheme. This allows for a direct comparison between the experimental data and theoretical predictions, without any amplitude fitting parameter. 


\subsection{Experimental results}
The blue curves in Figs.~\ref{fig:1}(b) and~\ref{fig:1}(c) are typical experimental curves for the magnitude $|I_D|$ and the phase $\phi_D = \mathrm{arg} \{I_D\}$ of the demodulated signal, at a vapor temperature of 53~$^\circ$C, corresponding to $b_0=75$ (other spectra at different temperatures are plotted in \ref{sec:exptsignal}). The red curves are the theoretical predictions that take into account the hyperfine structures of the excited state and Doppler broadening, but leave out the Zeeman manifold (see~\ref{sec:modelTrans} for the complete derivation). The theoretical curves capture well the qualitative behavior of the experimental signals. Far away from the spectrum center, we observe a small frequency shift in the spectroscopic features, between theory and experimental data. This could be due to a slight nonlinearity in the scan of the laser frequency that is not captured by a linear calibration of the frequency axis. Residual amplitude modulation (RAM) of the probe beam, which modifies the sideband spectrum, can result from the modulator. The RAM is known to affect the modulation transfer spectroscopic technique~\cite{jaatinen2008,preuschoff2018}. For our setup, however, we checked theoretically that the RAM level induced by our EOM does not significantly alter the spectroscopic signals, and can be disregarded in our analysis.

To understand the key characteristics of those spectra, we show in \fig{1}(d) the expected signal for a two-level medium, calculated at the same density and temperature of Figs.~\ref{fig:1}(b) and~\ref{fig:1}(c). Its behavior is similar to the demodulated signal observed for the cesium D2 line, indicating that the hyperfine structure does not play a major role in the overall structure of the spectra.  However, due to an exact cancellation of the contribution from the negative and the positive sidebands, the signal drops to zero for the two-level case at the spectrum center $\Delta_c = 0$. Since the in-phase and in-quadrature components are anti-symmetric in detuning $\Delta$ (see~\ref{sec:generaltheory}), the phase of the demodulated signal experiences an abrupt $\pi$ shift at resonance. 

When several atomic transitions contribute to FM spectroscopy signal, such as in the cesium D2 line, the spectrum becomes asymmetric and there is no more exact cancellation of the contributions of the negative and positive sidebands. Nevertheless, the magnitude of the demodulated signal still exhibit a minimum that we take as the spectrum center $\Delta_c$ [black dashed line in Figs.~\ref{fig:1}(b) and~\ref{fig:1}(c)]. $\Delta_c$ also coincides with a rapid change of the phase by $\pi$, as for the two-level case.

A striking feature of the amplitude spectrum is its narrow peak at the spectrum center. Since the Doppler broadening {\it rms} value is about 30$\Gamma$, this narrow peak is clearly sub-Doppler. Furthermore, this peak becomes narrower as the OD increases, as shown in the plot of the demodulated component $I'_p$ for several ODs in \fig{1}(e). This component is defined by $I'_P=\mathrm{Re}\left\{I_D\mathrm{e}^{i\varphi}\right\}$, where $\varphi=-\mathrm{arg}\left\{dI_D/d\Delta|_{\Delta=\Delta_c}\right\}$. Physically, by applying a phase rotation of $\varphi$, we transfer fully the slope at $\Delta_c$ of the demodulated signal to the component $I'_P$. Consequently, the component $I'_Q$ which is in quadrature to $I'_P$,  has a slope $dI'_Q/d\Delta|_{\Delta=\Delta_c}=0$. $I'_P$ shows a dispersive-like behavior at the vicinity of $\Delta_c$, similar to the usual FM spectroscopy technique \cite{Bjorklund:80}. Since the dominant sidebands of the probe laser are off-resonance, and explore the slow decay tails of the absorption window, this narrow structure could not come from the absorptive response of the atomic vapour. They rather come from the rapid variation of the phase of the first sidebands as they propagate through the medium. This phase variation increases with the OD leading to the sub-Doppler structures at large OD, as observed in Figs.~\ref{fig:1}(b)--\ref{fig:1}(e).

\begin{figure}[htbp]
   \begin{center}
     \includegraphics[width =0.75\linewidth]{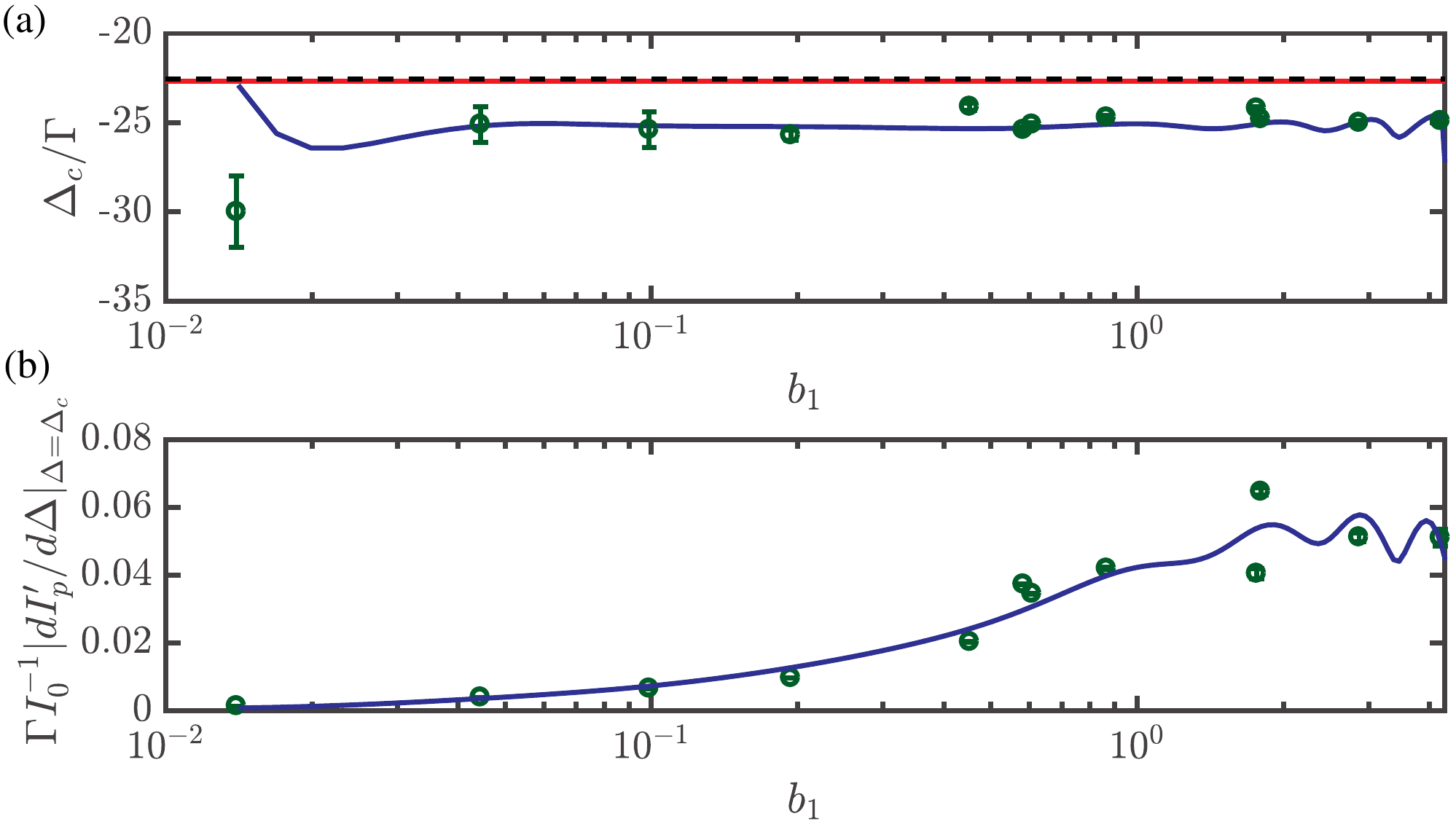}
      \caption{(a) The center of the cesium D2 line, and (b) the magnitude of the slope at the spectrum center plotted against $b_1$, which is the average OD of the first sidebands, when the carrier is tuned to the line center. The blue curves are the theoretical predictions. The green open circles are the experimental values. In (a), the black dashed line indicates the geometrical center of the three allowed transitions. The red curve is calculated for an OD that is ten times larger, with a larger modulation frequency of 2~GHz. The error bars represent the statistical errors of one standard deviation computed during the fitting procedure. Hence, the error bars do not take into account the shot to shot fluctuations in the experiment.}
      \label{fig:2}
   \end{center}
\end{figure}

In \fig{2}(a), the spectrum centers $\Delta_c$, measured at various temperatures are shown as green open circles. Due to the excited state hyperfine structure, $\Delta_c$ does not coincide with the $F=4\rightarrow F'=5$ transition, for which $\Delta = 0$. The horizontal axis variable 
\begin{equation}
b_1=b_0\Gamma^2/(4\Omega^2)
\label{eq:b_1}
\end{equation}
is the OD of the first sideband when $\Delta= \Delta_c$~\bibnote{In the experiment, the spectrum is asymmetric, so $b_1$ corresponds to the averaged value of the ODs at the $+1$ and $-1$ sidebands, when $\Delta= \Delta_c$}. The experimental data are in good agreement with the theoretically calculated value (blue curve). We note that the value of $\Delta_c$ varies for small and large value of $b_1$ ($b_1<0.05$ and $b_1>1$ in this case), which might prevent us to use this medium for accurate frequency reference. Moreover, the value of $\Delta_c$ does not correspond to any physical relevant quantity of the system, since it results from a subtle balance between the contribution of the positive and negative sidebands on the asymmetric spectrum. In contrast, for larger modulation frequency such that the excited state hyperfine splitting becomes negligible with respect to $\Omega$, the center value becomes independent of $b_1$ [see the red curve in \fig{2}(a)]. In this situation, the spectrum center has a clear physical meaning; it corresponds to the geometrical center defined as  $\sum_iS_i\Delta_i/\sum_iS_i$, where $S_i$ is the transition strength factor and $\Delta_i$ is the frequency splitting of the hyperfine excited state Zeeman manifold $i$ [see dashed line in \fig{2}(a)].

The dimensionless maximal slope of the demodulated signal at the spectrum center,\linebreak $\Gamma I_0^{-1}|dI'_{P}/d\Delta|_{\Delta=\Delta_c}$ is shown in \fig{2}(b). This slope is used as a figure-of-merit for the frequency sensitivity of the spectroscopic method. The experimentally measured values of the slope [see green open circles in \fig{2}(b)] are in good agreement with the calculated ones (blue curve). The sensitivity increases with $b_1$ and reaches a maximum value of $\sim0.05$ for $b_1\simeq 2$. For media with higher OD, the sensitivity is expected to decrease due to an increase in the absorption of the first sidebands that leads to an overall reduction of the transmitted signal. Nevertheless, according to \eq{b_1}, one can increase the modulation frequency to prevent a large value of $b_1$. In this context, we can show numerically that the sensitivity can be further increased. 

\section{Discussions}
Now, we discuss the frequency sensitivity of the large OD FM spectroscopic technique, more precisely, on how the slope at spectrum center depends on experimental parameters. As shown in \fig{1}, the main spectroscopic features are well captured by a two-level medium. Hence, for the sake of simplicity, we center our discussions only on a two-level medium.

We first consider the large OD FM spectroscopy applied to a two-level medium at $T=0$. In the limit of $\Omega\gg\Gamma$ that brings the sidebands into the tail of the resonance, the following expression is found for the slope at $\Delta_c$ (see details of the derivation in~\ref{sec:highmod}), 
\begin{equation}
\Gamma I_0^{-1}|dI'_{P}/d\Delta|_{\Delta=\Delta_c}\approx \frac{3}{2}J_1(\beta)J_2(\beta)b_1\exp\left(-\frac{5}{8}b_1\right).
\label{eq:Dev_High}
\end{equation}
where $J_n(x)$ is the $n$-th order Bessel function of the first kind. We consider only the first and second sidebands, the others are supposed to be too weak or too detuned to give a noticeable contribution. A maximal sensitivity of $\sim 0.2$ is obtained for $b_1=8/5$, and $\beta=2.4$. We note that the experiment [see \fig{2}(b)] gives a sensitivity around 4 times smaller than the prediction of  \eq{Dev_High}. This lower value is due to the residual effects of Doppler broadening and hyperfine structure in the experiment. The lower value of $\beta=2.14$ used in our experiment, leads only to a 4\% reduction in the sensitivity of the spectroscopic technique. Numerical simulations show that the maximal sensitivity is obtained when $b_0\gtrsim 2000$, which is about 10 times larger than the experimental maximal OD.

Considering now the usual low-modulation-index FM spectroscopy at $T=0$ \cite{Bjorklund:80}, the sensitivity is found to be (see also~\ref{sec:lowmod})
\begin{equation}
\Gamma I_0^{-1}|dI'_{P}/d\Delta|_{\Delta=\Delta_c} \approx 2J_0(\beta)J_1(\beta)b_0\exp\left(-\frac{b_0}{2}\right),
\label{eq:Dev_Low}
\end{equation}
where we consider only the carrier and the first sidebands. A maximum sensitivity of $0.5$ is found for $b_0=2$ and $\beta=1$, which is  larger but comparable to the high index  case [see \eq{Dev_High}]. Note that there is an optimum OD of $b_0=2$ for the low modulation index case, unlike in the high modulation index case, where there is no OD limitation for optimum sensitivity, as long as $\Omega$ can be adjusted to have $b_1\simeq 8/5$ [see \eqs{b_1}{Dev_High}].

\begin{figure}[htbp]
   \begin{center}
      \includegraphics[width =0.75\linewidth]{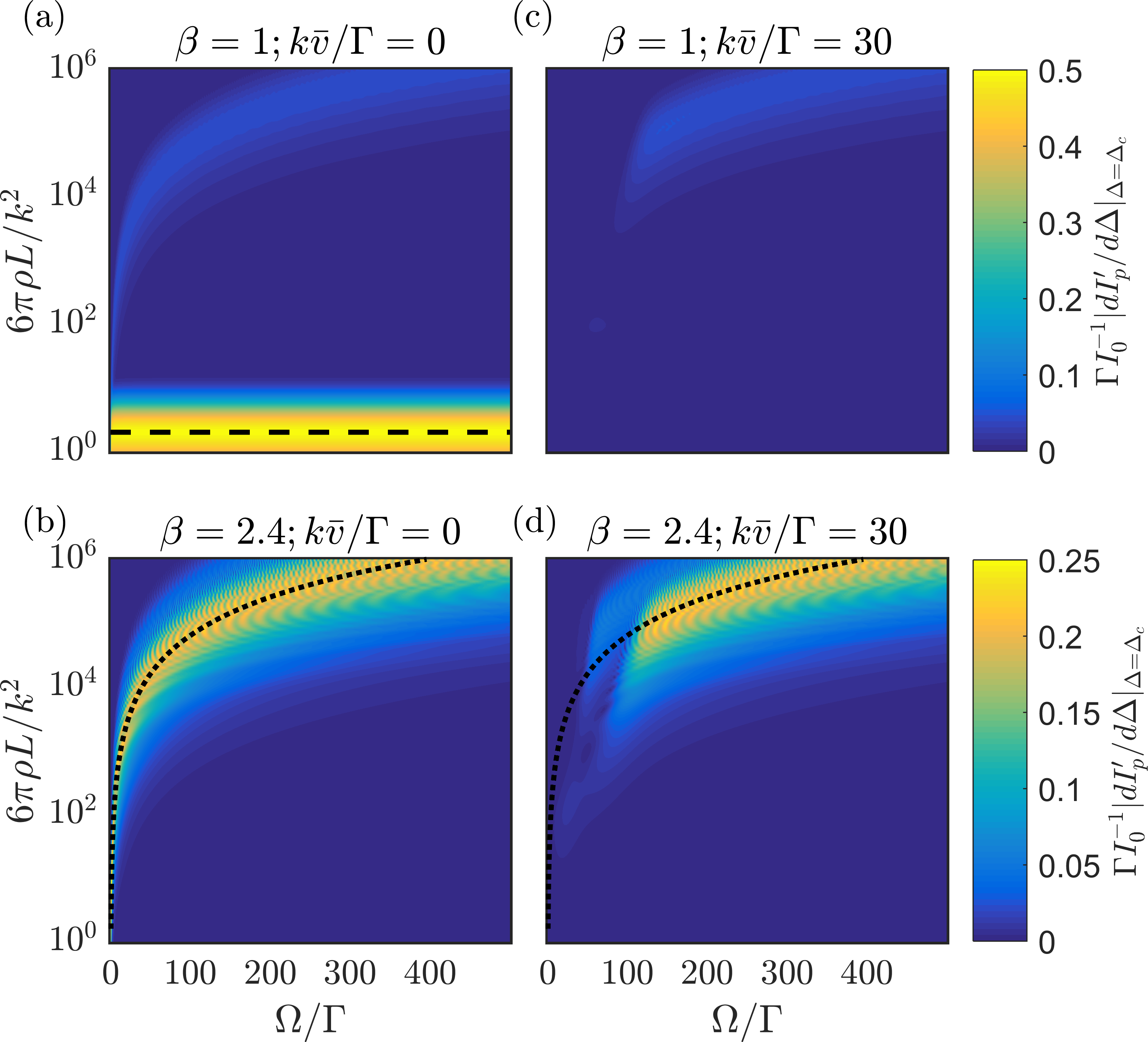}
      \caption{2D maps of the calculated sensitivities of FM spectroscopy applied to a two-level system with (a) $\beta = 1$ at  $k\vbar/\Gamma=0$, (b) $\beta = 2.4$ at at  $k\vbar/\Gamma=0$, (c) $\beta = 1$ at $k\vbar/\Gamma=30$, and (d) $\beta = 2.4$ at $k\vbar/\Gamma = 30$.  The dashed line in (a), at $b_0=2$, identifies the maximal sensitivity in the low modulation index case. In (b) and (d), the maximal sensitivity at high modulation index of \eq{Dev_High}, is indicated by the dotted lines.}
      \label{fig:3}
   \end{center}
\end{figure}

A more complete numerical comparison of the sensitivities for the low and high modulation index cases is presented in Figs.~\ref{fig:3}(a) and~\ref{fig:3}(b) in the form of 2D maps. Here, we consider a two-level medium at $T=0$, and include all the possible relevant sidebands. We plot on the vertical axes the quantity $6\pi\rho L/k^2$, which corresponds to $b_0$ at $T=0$. $\rho$ is the atomic density and $k$ is the optical field wavenumber. The expressions of the sensitivity given by \eq{Dev_High} and \eq{Dev_Low} are represented by dotted and dashed curves, respectively. We note that those expressions capture well the position and the value of the maximum sensitivity. In Figs.~\ref{fig:3}(c) and~\ref{fig:3}(d), we extend the comparison to the finite temperature case. We consider a medium with a Doppler width of $k\vbar/\Gamma=30$, similar to our experiment. Here, $\vbar=\sqrt{k_BT/m}$ is the thermal velocity, $k_B$ is the Boltzmann constant and $m$ is the atomic mass. At $T\neq0$, the sensitivity of the standard low OD FM spectroscopy is reduced by Doppler broadening [compare Figs.~\ref{fig:3}(a) and~\ref{fig:3}(c) in the region where $6\pi\rho L/k^2\simeq2$]. In contrast, the maximal sensitivity of the high index FM spectroscopy, for sufficiently large $\Omega$, is still given by \eq{Dev_High}. This is shown in Fig.~\ref{fig:3}(d), where the full sensitivity of the $T=0$ case is recovered when $\Omega> 150\Gamma$. Here, $\Omega \gg k\vbar$, so the sidebands probe the tails of the resonance that are dominated by the homogeneous line rather than the Doppler broadening. Thus, the relevant parameter to compare the two temperature cases is indeed $6\pi\rho L/k^2$; the OD at $T=0$. We note that for finite temperature, we get
$b_0=6\pi\rho L g(k\vbar/\Gamma) / k^2$ where $g(x) = \sqrt{\pi/8}\exp\left(1/8x^2\right)\erfc\left(1/\sqrt{8}x\right)/x$~\cite{chalony_coherent_2011}. For large $x$, $g(x)\simeq\sqrt{\pi/8}/x$, leading to a substantial reduction of the OD (of a factor $\backsim k\vbar/\Gamma$) for the finite temperature medium compared to the $T=0$ case. 

In Figs.~\ref{fig:3}(a) and~\ref{fig:3}(c), a signal is also present at large OD. Indeed, at $\beta = 1$, the second sidebands of the modulation is not negligibly small, as $J_2(\beta=1) = 0.11$. Thus, while the carrier component is absorbed at large OD, the second sidebands start to probe the tails of the resonance, giving rise to a beat note with the first sidebands. Here, we have again the large OD FM spectroscopic technique, but operating away from $\beta =2.4$ where the sensitivity is optimum.

\section{Conclusion}
In conclusion, we presented a sensitive FM spectroscopic technique that uses the detuned sidebands to probe a large OD medium. When the modulation frequency becomes much larger than the Doppler width, these sidebands probe the tails of the resonance, which are dominated by the homogeneous response of the vapor. This leads to a Doppler-free technique with high sensitivity at large OD. Applying the large OD FM spectroscopy on the cesium D2 line, we find a good agreement with the calculated signal.  Applications might be found in measurement of cooperative emissions in dense atomic bulk medium where the spurious finite size effects shall be weak. 
Finally, this technique should be applicable to other types of media with large OD, such as dye or other molecular solutions, Mie scatterers ensemble, point-defects in diamond, and heavily doped glasses and crystals.

\appendix

\section{General expression for the demodulated signals}\label{sec:generaltheory}

We consider an incident field of amplitude $E_0$ that is phase modulated at a frequency $\Omega$,
\begin{equation}
E_i(t) = E_0 \mathrm{e}^{-i\omega t + i\beta \cos\Omega t}.
\end{equation}
The laser frequency is denoted by $\omega$, and the modulation index for the phase is denoted by $\beta$. Using the Jacobi-Anger expansion and the relation  $J_{-n}(x) = (-1)^n J_n(x)$, we have
\begin{equation}
E_i(t)=E_0\sum_{n=-\infty}^{\infty}i^nJ_n(\beta)\mathrm{e}^{-i\left(\omega+n\Omega\right)t}.
\label{eq:Ei}
\end{equation}

The transmitted field across a homogeneous medium is given by 
\begin{equation}
E(t) = E_0 \sum_{n=-\infty}^\infty i^n J_n(\beta) \mathrm{e}^{-i(\omega+ n\Omega) t}B_n(\Delta),
\end{equation}
where the function $B_n(\Delta)$ is the transmittivity of the $n$-th sideband, and $\Delta$ is the detuning of the carrier frequency. Under the condition that we do not saturate the atomic transition, the transmittivity for a medium of thickness $L$ is given by
\begin{equation}
B_n(\Delta) = \exp[i\chi(\Delta +n\Omega) kL/2],\label{eq:transmittivity}
\end{equation}
where $\chi(\Delta)$ is the susceptibility of the medium. As a result of the frequency modulation, the transmitted intensity consists of various harmonics of $\Omega$:
\begin{equation}
{I(\Delta, t)} = I_0 \sum_{n,m} i^{n-m}J_n(\beta) J_m(\beta) \mathrm{e}^{-i(n-m)\Omega t} B_n(\Delta) B_m^*(\Delta),
\end{equation}
where $n$ and $m$ are summed over all integers. 


We are  interested in the first harmonic of transmitted intensity
\begin{align}
I_1(\Delta, t) =& i I_0\sum_{n=-\infty}^\infty\Big[J_n(\beta)J_{n-1}(\beta) B_n(\Delta)B_{n-1}^*(\Delta) \mathrm{e}^{-i\Omega t} - J_n(\beta) J_{n+1}(\beta) B_n(\Delta) B_{n+1}^*(\Delta)\mathrm{e}^{i\Omega t}\Big]\nonumber\\
=& 2I_0 \sum_{n=-\infty}^\infty J_n(\beta)J_{n+1}(\beta) \Big[\im \left\{B_n(\Delta)B_{n+1}^*(\Delta)\right\}\cos \Omega t + \re \left\{B_n(\Delta)B_{n+1}^*(\Delta)\right\}\sin \Omega t\Big]
\end{align}
Applying the relation $J_{-n}(x) = (-1)^n J_n(x)$, we can rewrite the expression above such that the summation is only over positive integers,
\begin{align}
I_1(\Delta, t) =& 2I_0 \sum_{n=0}^{\infty} J_{n}(\beta)J_{n+1}(\beta) \Big[\mathrm{Im}\{B_n(\Delta)B_{n+1}^*(\Delta)-B_{-n}^*(\Delta)B_{-n-1}(\Delta)\}\cos\Omega t\nonumber\\
&+ \mathrm{Re}\{B_n(\Delta)B_{n+1}^*(\Delta)-B_{-n}^*(\Delta)B_{-n-1}(\Delta)\}\sin\Omega t\Big].
\end{align}
The in-phase and in-quadrature time-averaged components of the demodulated signal are related to $I_1$ through $I_1= 2(I_P\cos\Omega t + I_Q \sin \Omega t)$. We identify those components as
\begin{align}
I_P(\Delta) & =  I_0\sum_{n=0}^{+\infty}J_n(\beta)J_{n+1}(\beta) \im\Big\{B_n(\Delta)B^*_{n+1}(\Delta)-B^*_{-n}(\Delta)B_{-n-1}(\Delta)\Big\},\nonumber\\
I_Q(\Delta) & = I_0\sum_{n=0}^{+\infty}J_n(\beta)J_{n+1}(\beta) \re\Big\{B_n(\Delta)B^*_{n+1}(\Delta)-B^*_{-n}(\Delta)B_{-n-1}(\Delta)\Big\}.\label{eq:IPQ}
\end{align}
We can also express the signal using the complex notation,
\begin{equation}
I_D(\Delta)\equiv I_P + i I_Q = iI_0\sum_{n=0}^{+\infty}J_n(\beta)J_{n+1}(\beta) \left\{B^*_n(\Delta)B_{n+1}(\Delta)-B_{-n}(\Delta)B^*_{-n-1}(\Delta)\right\}.\label{eq:Id}
\end{equation}
Under this notation, the first harmonic intensity can be written as
\begin{equation}
I_1(\Delta,t) = 2\mathrm{Re} \left\{I_D(\Delta) \mathrm{e}^{-i\Omega t}\right\}.
\end{equation}
For a given phase factor $\varphi$, we also have
\begin{equation}
I_1 = 2\mathrm{Re} \left\{I_D\mathrm{e}^{i\varphi} \mathrm{e}^{-i(\Omega t+\varphi)}\right\}.
\end{equation}
Thus, a change in the phase of the demodulation reference signal by $\varphi$, results in a demodulated signal that is rotated by $\varphi$ in the complex plane, \ie,
\begin{equation}
I_D' = I_D \mathrm{e}^{i\varphi}.\label{eq:rotatedId}
\end{equation}

For the simple case of a zero-temperature ($T=0$) two-level medium with a density $\rho$, the susceptibility is given by
\begin{equation}
\chi(\Delta)=-3\pi\rho\Gamma/\left[k^3(\Delta+i\Gamma/2)\right].
\end{equation}
We define
\begin{equation}
b = \im\{\chi\}kL,
\end{equation}
as the OD, and,
\begin{equation}
\phi = \re\{\chi\} kL/2,
\end{equation}
as the optical phase shift due to the refractive index of the two-level ensemble. At $T=0$, the OD at resonance $b_0$ is given by
\begin{equation}
b_0 = \frac{6 \pi \rho L}{k^2}.
\end{equation}

The symmetric property of the susceptibility gives rise to the following relation for the transmittivity,
\begin{equation}
B_n(\Delta) = B_{-n}^*(-\Delta).
\end{equation}
Thus, for the two-level case, we find that the in-phase and in-quadrature components are both odd functions of $\Delta$. As pointed out before, this leads to an abrupt phase jump of $\pi$ across the resonance. 

We further note that, in the two-level case, the spectrum center $\Delta_c$ occurs at resonance \ie, $\Delta_c=0$.

\section{High modulation index case}\label{sec:highmod}
We consider here the high modulation index case, which forms the basis for the large OD frequency modulation (FM) spectroscopy. We suppose that the carrier component is weak and the signal is dominated by the beat note between the 1st and the 2nd sidebands. The in-phase and in-quadrature components simplifies to the following:
\begin{align}
I_P(\Delta) & =  I_0 J_1J_{2} \im\left\{B_1(\Delta)B^*_{2}(\Delta)-B^*_{-1}(\Delta)B_{-2}(\Delta)\right\},\nonumber\\
I_Q(\Delta) & = I_0 J_1J_{2} \re\left\{B_1(\Delta)B^*_{2}(\Delta)-B^*_{-1}(\Delta)B_{-2}(\Delta)\right\}.
\end{align}
In the complex notation, we have
\begin{equation}
I_D(\Delta) = iI_0 J_1J_{2} \left\{B^*_1(\Delta)B_2(\Delta)-B_{-1}(\Delta)B^*_{-2}(\Delta)\right\}.
\end{equation}

We further assume that the modulation frequency is sufficiently large, \textit{i.e.} $\Omega\gg\Gamma$, so that the two-level susceptability can be approximated by
\begin{equation}
\chi(\Delta) \approx -\frac{3\pi\rho\Gamma}{k^3}\left(\frac{1}{\Delta} - \frac{i\Gamma}{2\Delta^2}\right).\label{eq:tail2level}
\end{equation}
For a medium with non-zero temperature, the above approximation also holds as long as the modulation frequency is much larger than the Doppler broadening.

Using~\eq{transmittivity} and~\eq{tail2level}, we can write
\begin{align}
B_{\pm1}(\Delta) \approx& \exp\left\{-\frac{b_0}{2}\left(\frac{\Gamma^2}{4(\Delta\pm\Omega)^2} +i \frac{\Gamma}{2(\Delta\pm\Omega)}\right)\right\},\nonumber\\
B_{\pm2}(\Delta) \approx& \exp\left\{-\frac{b_0}{2}\left(\frac{\Gamma^2}{4(\Delta\pm2\Omega)^2} +i \frac{\Gamma}{2(\Delta\pm2\Omega)}\right)\right\}.
\end{align}
The product of the transmittivity function in \eq{Id} can be written as
\begin{multline}
B_{\pm1}B^*_{\pm2} \approx \exp\left\{-\frac{b_0}{2}\frac{\Gamma^2}{4(\Delta\pm\Omega)^2}\right\}\exp\left\{-\frac{b_0}{2}\frac{\Gamma^2}{4(\Delta\pm2\Omega)^2}\right\}\\
\times\exp\left\{-i\frac{b_0}{2}\frac{\Gamma}{2(\Delta\pm\Omega)}\right\}\exp\left\{i\frac{b_0}{2}\frac{\Gamma}{2(\Delta\pm2\Omega)}\right\}
\end{multline}
Its derivative with respect to $\Delta$, and subsequent evaluation at the spectrum center ($\Delta = 0$), is given by
\begin{equation}
\frac{\ud }{\ud \Delta }B_{\pm1}B^*_{\pm2}|_{\Delta =0} \approx iB_{\pm1}(0)B_{\pm2}^*(0) \frac{b_0}{\Omega}\frac{3\Gamma}{16\Omega}.
\end{equation}
Since $\Omega\gg\Gamma$,  we retain only the first order terms in $\Gamma/\Omega$. Therefore, we have
\begin{equation}
\left.\frac{\ud }{\ud \Delta }\left[B_{1}^*B_{2}-B_{-1}B_{-2}^*\right]\right|_{\Delta = 0} \approx -\frac{3i}{2} b_1 \exp\left({-5b_1/8+i\phi_1/2}\right)\Gamma^{-1}.
\end{equation}
We denote the OD and the optical phase shift at the position of the first sidebands to be $b_1 =b_0\Gamma^2/(4\Omega^2)$ and $\phi_1=b_0\Gamma/(4\Omega)$ respectively, when the carrier component is at the center of the spectrum. The slope of the demodulated signal is then given by
\begin{equation}
\left.\frac{\ud I_D}{\ud \Delta}\right|_{\Delta = 0} \approx \frac{3}{2} I_0 J_1 J_2 b_1 \exp\left({-\frac{5}{8}b_1+\frac{i}{2}\phi_1}\right)\Gamma^{-1}.
\end{equation}
As the OD of the medium changes,  $\phi_1$ changes and the values of the slope for the in-phase and in-quadrature components display an oscillatory behavior. The sensitivity of the spectroscopic technique can be measured by a suitable phase shift of the reference signal, according to equation \eq{rotatedId}, such that the component $I_P'=\mathrm{Re}\{I_D'\}$, has the full value of the on-resonance slope while $I_Q'=\mathrm{Im}\{I_D'\}$ has zero slope. In practice, it is a measurement of the magnitude of the slope, given by
\begin{equation}
|\ud I_P'/\ud \Delta|_{\Delta=0} \approx \frac{3}{2} I_0J_1 J_2 b_1 \exp\left({-\frac{5}{8}b_1}\right)\Gamma^{-1},
\end{equation}
which is \eq{Dev_High}. The maximum value of $J_1(\beta)J_2(\beta)$ occurs when $\beta=2.4$, giving the optimum  modulation index for large OD FM spectroscopy.

\section{Low modulation index case}\label{sec:lowmod}
We contrast the results obtained in the previous section with the case of low modulation index. For low modulation index, one only has to consider the beat note between the carrier and the first sidebands.  The demodulated signal becomes
\begin{equation}
I_D(\Delta) =iI_0 J_0J_{1} \left\{B^*_0(\Delta)B_1(\Delta)-B_{0}(\Delta)B^*_{-1}(\Delta)\right\}.\label{eq:generalId}
\end{equation}
When $\Omega\gg\Gamma$, we have the conventional band-resolved FM spectroscopy 

In the limit of low OD ($b_0\ll 1$), we can approximate $B_{\pm1} \approx 1$ and $B_0 \approx 1 - b_0/2 + i\phi$. The demodulated signal becomes
\begin{align}
I_P(\Delta) & = 2 I_0 J_0J_{1} \phi,\nonumber\\
I_Q(\Delta) & = 0.\label{eq:lowODslope}
\end{align}
The demodulated signal is non-zero only for the in-phase component. Furthermore, it has a dispersive profile suitable to generate an error signal for the frequency stabilization of a laser.

To compute the slope of the demodulated signals, we first note that
\begin{align}
B_0^*(\Delta)B_1(\Delta)-B_{0}(\Delta)B^*_{-1}(\Delta) \approx \left[B_0^*(\Delta)-B_0(\Delta)\right].
\end{align}
Its derivative, evaluated at the center, is then given by
\begin{equation}
\left.\frac{\ud }{\ud \Delta}\left[B^*_0(\Delta)-B_0(\Delta)\right]\right|_{\Delta=0}  \approx \frac{2i}{\Gamma}b_0 \mathrm{e}^{-b_0/2}.
\end{equation}

Here, the component $I_P'$ that has the full slope is simply $I_P$. The on-resonance slope is given by
\begin{equation}
|\ud I_P'/\ud \Delta|_{\Delta=0} \approx 2 I_0 J_0 J_1 b_0\mathrm{e}^{-b_0/2} \Gamma^{-1},
\end{equation}
which is \eq{Dev_Low}. Here, the maximum value $J_0(\beta)J_1(\beta)$ is obtained when $\beta = 1$.

\section{Experimental demodulated signals}\label{sec:exptsignal}
\begin{figure*}[!h]
\includegraphics[width = \textwidth]{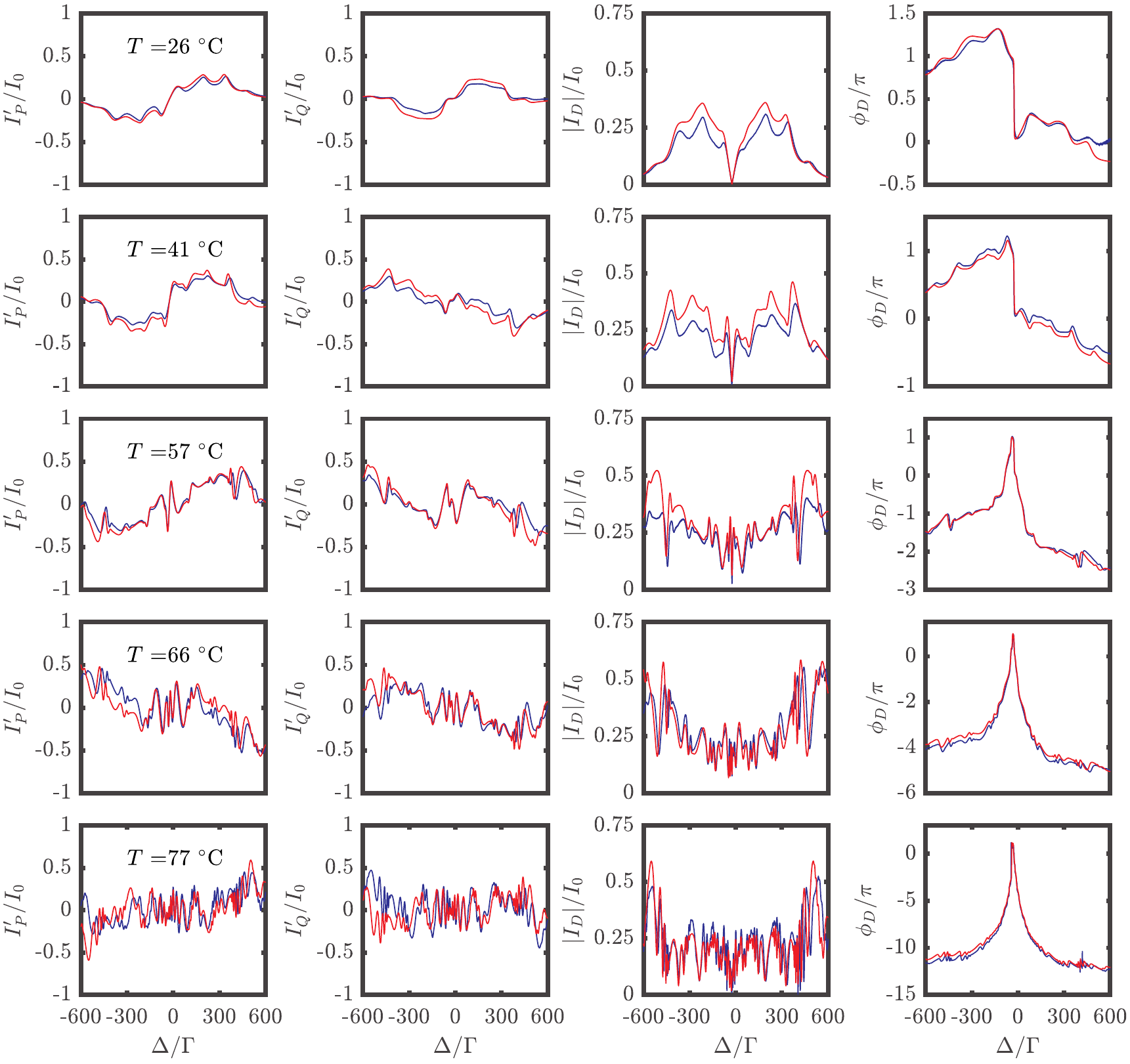}
\caption{The demodulated signals at several vapor temperatures. The first column and the second column show the $I_P'$ and the $I_Q'$ components. The third column and the fourth column  show the magnitude $|I_D|$, and phase of the demodulated signal, $\phi_D$. The blue curves are the experimental results and the red curves are the theoretical predictions including the three allowed transitions of cesium D2 line from the $F=4$ ground state. The vapor temperatures indicated here are the temperatures obtained from a fit of the theoretical model to the experimental curves. The fit is performed in a frequency range of 400 $\Gamma$ around $\Delta_c$. They agree well with direct measurements of the temperatures on the setup.}\label{fig:supfig}
\end{figure*}

Experimental demodulated signals at various vapor temperature are shown in \fig{supfig}. The experimental curves are plotted in blue, while the theoretical curves are plotted in red. In the first two columns, we plot the $I_P'$ and $I_Q'$ components of the demodulated signals. In the third and the fourth columns, we plot the magnitude $|I_D|$ and the phase $\phi_D=\arg\{I_D\}$ of the demodulated signals. As the vapor temperature increases, the demodulated signals become more complicated, as evidenced by the increasing oscillations in the magnitude, and the rapid change in the phase of the demodulated signals. 

\section{Model for the transmittivity of cesium D2 line}\label{sec:modelTrans}
To capture properly the contribution of the three-allowed transitions in a cesium vapor of temperature $T$ and thermal velocity $\vbar$, we use the following expression of the transmittivity at the vicinity of the D2 line
\begin{equation}
B(\Delta) = \exp\left[-\frac{\mathcal{B}}{2}\sqrt{\frac{\pi}{8}}\frac{\Gamma}{k\vbar}\sum_{F'=3}^5S_{4F'}w\left(\frac{\Delta-\delta_{F'}+i\Gamma/2}{\sqrt{2}k\vbar}\right)\right],\label{eq:transmittivity3}
\end{equation}
where $S_{FF'}$ is the transition strength factor.  They take the values $S_{4F'}=7/72$, 7/24 and 11/18, for $F'=3$, 4 and 5 respectively~\cite{steckCesium}. The detuning $\Delta$ is referred from the $F=4 \rightarrow F'=5$ transition. The two other relevant hyperfine excited states are detuned from the $F'=5$ level by $\delta_{F'}$. In this case, $\delta_{F'} = -452.4$, $-251.1$ and 0~MHz, for $F'=3$, 4 and 5 respectively~\cite{steckCesium}. The function $w(z)$, with a complex parameter $z$, is the Faddeeva function. It is defined by $w(z) = \exp(-z^2)\mathrm{erfc}(-iz)$~\cite{abramowitz}. $\mathcal{B}$ is a parameter proportional to $\rho L$, which is described in the following. We assume that the intensity of each sideband is low enough such that transition saturation and optical pumping can be neglected. We also neglect the contribution of the other $F=3$ hyperfine ground state, since it is 9.2~GHz away from the $F=4$ ground state. This transmittivity function is used in \eq{Id} to calculate the demodulated signals. 

The expressions of the absorption cross sections for the D lines of alkali atoms, are found in~\cite{siddons1}. Using the expression for the D2 line, we can write $\mathcal{B}$ in terms of the atomic density $\rho$,
\begin{equation}
\mathcal{B}= \frac{18\pi L \rho}{(2I+1) k^2},\label{eq:b0}
\end{equation}
where $I=7/2$ is the nuclear spin of cesium atoms.

The atomic density is then related to the vapor pressure $P_v$ and vapor temperature $T$, 
\begin{equation}
\rho = 133.323 \frac{P_v}{k_BT}.\label{eq:densityT}
\end{equation}
In the above expression, $T$ is specified in Kelvin and $P_v$ in Torr. The vapor pressure of cesium is further related to its temperature~\cite{vaporpressure},
\begin{align}
\log_{10} \frac{P_v}{760 \text{ {T}orr}} =  4.711 - \frac{3999\, \mathrm{K}}{T}, &\quad T < 301.64\, \mathrm{ K},\nonumber\\
\log_{10} \frac{P_v}{760 \text{ {T}orr}} =  4.165 - \frac{3830\, \mathrm{K}}{T}, &\quad T > 301.64\, \mathrm{ K}.\label{eq:PvT}
\end{align}
\eqs{densityT}{PvT} together link the temperature to $\mathcal{B}$. Thus, $\mathcal{B}$ and $T$ are not independent in our model here. Between the two quantities, we choose $T$ as the free parameter when fitting our experimental data with the model.

\section*{Funding}
Centre for Quantum Technologies (R-710-000-029-135); Singapore Ministry of Education (MOE2016-T3-1-006(S), MOE2018-T1-001-027).

\section*{Acknowledgments}
The authors wish to thank M. Ducloy, C. Monroe, C. Salomon, and N. I. Zheludev for fruitful discussions. Rustem Shakhmuratov acknowledges support from the FRC "Kazan Scientific Center of the Russian Academy of Sciences" and the Government Program of Competitive Growth of KFU.\\


\bibliography{CFM}

\end{document}